# Topological spin-texture transitions in van der Waals magnets revealed by X-ray Fourier transform holography

Sourav Chowdhury[1*#], Soumyaranjan Dash[2*], Michael Schneider[3], Christopher Klose[3], Chithra H. Sharma[4,5], Lisa-Marie Kern[3], Tim A. Butcher[3,6], Josefin Fuchs[3], Santanu Pakhira[7,8], Samik DuttaGupta[9], Takashi Taniguchi[10], Kenji Watanabe[11], Sujit Das[12], Sanjeev Kumar[2#], Bastian Pfau[3], Amir-Abbas Haghighirad[7#], & Moritz Hoesch[1#]

[1]Deutsches Elektronen-Synchrotron DESY, Notkestrasse 85, 22607 Hamburg, Germany.
[2]Indian Institute of Science Education and Research (IISER), Mohali, Sector 81, S.A.S. Nagar, 140306, Manauli, India.
[3]Max Born Institute for Nonlinear Optics and Short Pulse Spectroscopy 12489 Berlin, Germany.
[4]Christian-Albrechts-Universität zu Kiel, Leibnizstrasse 19, 24098 Kiel, Germany.
[5]Universität Hamburg, Luruper Chaussee 149, Hamburg 22761 Germany.
[6]Paul Scherrer Institut, 5232 Villigen PSI, Switzerland.
[7]Karlsruhe Institute of Technology, Kaiserstr. 12, D-76131 Karlsruhe, Germany.
[8]Department of Physics, Maulana Azad National Institute of Technology, Bhopal, M.P. - 462003, India.
[9]Saha Institute of Nuclear Physics, 1/AF Bidhannagar, Kolkata 700064, India.
[10]Research Center for Materials Nanoarchitectonics, National Institute for Materials Science, 1-1 Namiki, Tsukuba 305-0044, Japan.
[11]Research Center for Electronic and Optical Materials, National Institute for Materials Science, 1-1 Namiki, Tsukuba 305-0044, Japan.
[12]Materials Research Centre, Indian Institute of Science, 560012 Bengaluru, Karnataka, India.

*Authors contributed equally
#Corresponding authors: (SC) sourav.chowdhury@desy.de/mailatsourav07@gmail.com, (SK) sanjeev@iisermohali.ac.in, (A-AH) amir-abbas.haghighirad@kit.edu, (MH) moritz.hoesch@desy.de

**Nontrivial topological spin-textures, such as skyrmions, merons, bimerons, and skyrmioniums, are envisioned as robust building blocks for future memory and logic devices. Controllable transformations between these states require a quantum-mechanical description of electronic degrees of freedom and atomic-scale insight beyond existing phenomenological models. Here, we report an atomic-scale investigation of topological phase transitions and their protection in the two-dimensional van der Waals ferromagnet $Fe_3GeTe_2$ (FGT) using a combined experimental-theoretical approach. Synchrotron-based Fourier transform holography directly images labyrinth domains, isolated skyrmions, mixed labyrinth-skyrmion phases, and skyrmion bags with high spatial resolution. We compare these observations to simulations based on an electronic lattice Hamiltonian that captures both metallicity and relativistic spin-orbit coupling in FGT. By systematically exploring a broad range of temperatures and magnetic fields, we map the mechanisms governing topological transitions and their stability. This sequential-integrated experimental-theoretical framework advances understanding of spin-texture interactions and enables precise control of external tuning parameters. Our results establish a platform for creating, stabilizing, and manipulating topological states, paving the way for engineered spin-texture transitions in next-generation spintronic technologies.**

## Introduction

Topological nanoscale spin-textures are attracting widespread interest for spintronic memory and logic technologies because of their high-density data storage, energy-efficient operation, and potential for unconventional computation[1-3]. Magnetic skyrmions, in particular, combine chiral, particle-like spin-configurations with topological protection, making them promising candidates for next-generation devices[4-6]. Their nanometric size, robustness, and rapid response to external stimuli (temperature, magnetic field, and electric bias)[7-9] place them at the forefront of proposals for racetrack memory[10], logic[11], and emerging computing paradigms such as quantum and neuromorphic approaches[12-14].

Skyrmions were first observed in bulk 3D magnet MnSi[15], which spurred extensive studies of similar textures in other 3D materials and nanostructures[16,17]. However, integrating such bulk systems into scalable devices remains challenging due to fabrication and dimensional constraints. 2D van der Waals (vdW) magnets offer a complementary platform: atomically thin layers with robust magnetic order down to the monolayer, flexible interfacial tuning, and rich opportunities for heterostructures[18-22]. Topological textures have recently been reported in several 2D vdW ferromagnets, including $CrI_3$[1,2,20,23], $CrBr_3$[24], $CrGeTe_3$[25,26], $Fe_3GeTe_2$ (FGT)[7,8,27-30], $Fe_3GaTe_2$[31], and $Cr_{1+x}Te_2$[32,33]. Particularly, recent studies on $Fe_3GaTe_2$ have revealed a rich skyrmionic phase space, including distinct room-temperature skyrmion phases[34], sub-100 nm Néel-type skyrmions in non-stoichiometric $Fe_{3-x}GaTe_2$[35], and high-temperature Néel skyrmions stabilized by Fe intercalation into the vdW gap[36]. Among these, FGT has drawn particular attention for gate-tunable room-temperature ferromagnetism[13], pronounced anomalous Hall effects[37], tunable magnetic anisotropy via electric field and carrier density[22,38,39], and exchange-bias phenomena[40,41], which together enable access to a diverse landscape of topological states. Notably, FGT exhibits transitions between distinct topological states, providing a versatile platform to study their microscopic origins[7,8,42]. Yet, despite progress, the microscopic mechanisms driving transitions between topological phases remain poorly understood, in part because conventional micromagnetic models, based on phenomenological descriptions, treat spins as classical building blocks and neglect the response of the electronic fluid in metals, including itinerant electrons and spin-orbit coupling (SOC). Consequently, the link between real-space topology and the underlying quantum electronic fluid, and, in particular, how the electronic fluid shapes real-space topological-textures and its implications for device performance, remains largely unexplored.

FGT offers an attractive testbed because of its strong SOC and itinerant electrons, which couple intimately to magnetization textures. As a correlated Hund's metal[43], its electronic properties and spin-textures are entwined, and the response of skyrmion configurations to temperature and magnetic field may differ from that in insulators or uncorrelated metals. However, a faithful atomistic bridge between imaging of topological textures and quantum-mechanical modeling under varying conditions has been lacking, limiting predictive control over topological phase transitions. Developing such a framework is essential to uncover the microscopic physics of topological states and to enable engineered spin-texture transitions for spintronic applications.

High-resolution soft X-ray imaging directly visualizes spin-textures in an FGT thin flake. By varying temperature and magnetic field, we reveal labyrinthine domains, isolated skyrmions, and skyrmion bags. Monte Carlo simulations based on an atomistic Hamiltonian that includes itinerancy and relativistic SOC reproduce these textures, essential to FGT's metallic character and magnetism. An excellent agreement across temperatures and magnetic fields exposes the microscopic mechanisms governing topological transitions and their protection. This framework highlights electronic structure, metallicity, and SOC in stabilizing and switching topological states. Temperature- and field-dependent transitions reveal unusual dynamics from competing energy scales and entropy. Our work provides a platform for controlling spin-textures in 2D magnets for spintronics.

## Results

### Integrated experimental-theoretical approach.

To study topological transitions in FGT, we adopt an integrated approach (Fig. 1) that combines high-resolution imaging with atomistic simulations. High-resolution imaging of the magnetic textures in the FGT flake is achieved by synchrotron-based Fourier transform holography (FTH)[44-46] with numerical phase retrieval[47], reaching ~15 nm spatial resolution and surpassing previous scanning transmission X-ray microscopy (STXM) studies[7,8]. Compared with conventional laboratory-based magnetic imaging techniques such as Lorentz-TEM[34-36], MFM[24,35], and optical Kerr microscopy[24], FTH with numerical phase retrieval provides the important combination of high spatial resolution, full-field imaging, and rapid data acquisition. The integrated sample-holography-mask assembly (Fig. 1 and Supplementary Fig. 2) minimizes drift, allowing reliable imaging of the same microscopic region after temperature variation and during magnetic-field cycling. This capability is essential for resolving the evolution and transformation pathways of labyrinth domains, isolated skyrmions, mixed labyrinth-skyrmion phases, and skyrmion bags. Moreover, FTH does not require a conventional imaging lens, but instead exploits a holography mask together with the high coherence of synchrotron X-rays. To the best of our knowledge, this study represents the first application of FTH to a 2D vdW material, enabled by a specialized sample-preparation methodology (see Methods and Supplementary Fig. 2 for details).

Our sample is a thin FGT flake (~150 nm thick) exfoliated from a single crystal and transferred onto a 150 nm $Si_3N_4$ membrane. Details of sample preparation and characterization of elemental composition and magnetic properties are provided in Methods and Supplementary Figs. 1-2. On the membrane's opposite side, a metallic holography mask with a 1.9 µm circular aperture defines the field of view. Contrast for the out-of-plane component ($m_z$) of the magnetic texture is achieved via X-ray magnetic circular dichroism (XMCD) at the Fe $L_3$-edge (708 eV, 1.75 nm X-ray wavelength).

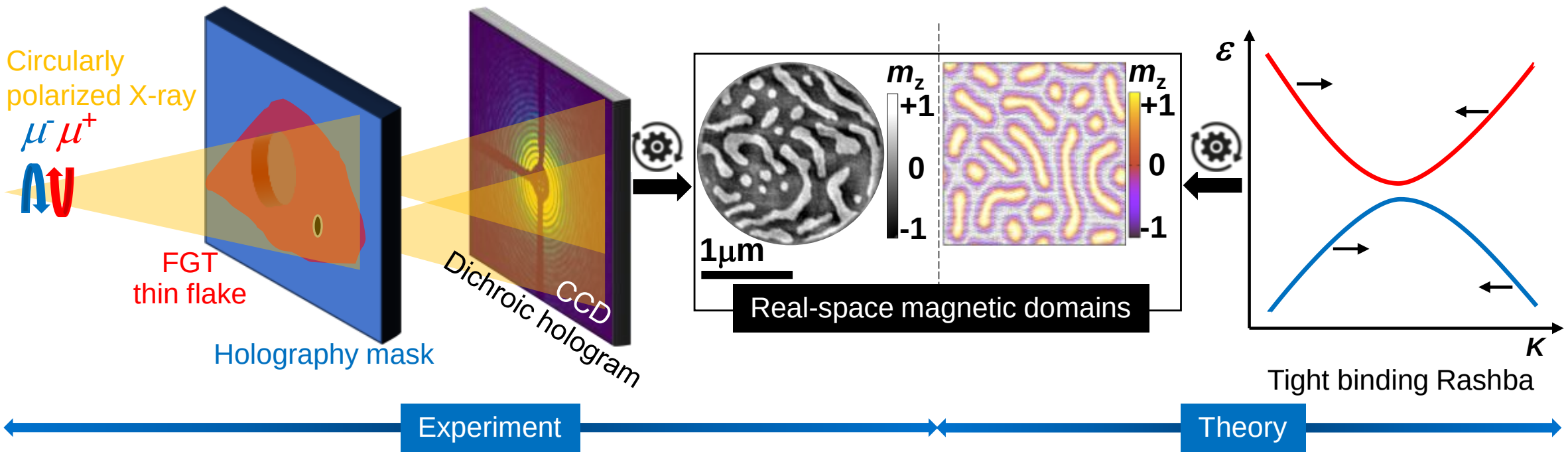

**Figure 1| Strategic synergy between experiment and theory. Experimental:** A circularly polarized X-ray beam from a synchrotron is incident on an FGT thin flake through a holography mask, producing an interference pattern recorded on a charge-coupled device (CCD) detector. XMCD holograms are processed to reconstruct real-space magnetic-textures. **Theoretical:** The corresponding magnetic nanodomain-textures are simulated with an atomistic tight-binding Rashba model that captures the electronic and magnetic behavior of the FGT system. Experimental data are shown in grayscale (circular); theoretical simulations are shown in color (rectangular). The color bars indicate the z-component of magnetization ($m_z$).

The experimentally observed nanodomain structures were directly compared with Monte Carlo simulations based on an electronic lattice Hamiltonian derived from a tight-binding Rashba model[48-53]. This model incorporates Hund's coupling, SOC, and uniaxial anisotropy to capture FGT's electronic and magnetic behavior. The simulations provide direct access to atomic-scale spin-configurations, enabling quantitative comparison of the experimental spin-textures. The theoretical framework begins with a fully quantum-mechanical lattice model describing the electronic degrees of freedom (Eq. 1, Methods). An effective spin model (Eq. 2, Methods) is then derived, incorporating ferromagnetic exchange, anisotropic Dzyaloshinskii-Moriya interactions (DMI), and a dipolar-like term, each arising from the interplay of SOC, $\lambda$, Hund's exchange, $J_H$, and uniaxial anisotropy, $A_u$. This model is simulated with a Markov-chain Monte Carlo algorithm (Methods), treating the temperature $T$ and the Zeeman field $H_Z$ as the controlling parameters. All energy scales are measured in units of the hopping parameter $t$ (detail in Method and Supplementary note 1). Based on band-structure calculations, FGT's effective bandwidth is ~2 eV[54,55], corresponding to a hopping parameter of 0.20-0.22 eV for a 2D triangular lattice, yielding Curie temperature ($T_C \approx 205$ K) in reasonable agreement with experiments. Mapping the theoretical magnetic-field scale to experimental values is nontrivial, as it depends on the effective g-factor of the magnetic moments, which SOC can strongly modify[56-59]. In the absence of precise g-factor data for FGT, we calibrate the field scale by matching simulated spin-patterns to the experimental field-cooling value of -8 mT.

To characterize the topological phases, we compute both the skyrmion density ($\chi_s$) and the Bott index ($B$) (see Methods and Supplementary Notes 1 and 2 for details). The density $\chi_s$ reflects the topology of the magnetization textures but does not capture the behavior of the underlying electronic fluid's response. In contrast, the Bott index $B$ characterizes the topology of the electronic fluid in the presence of magnetic inhomogeneities[60-66], arising from non-periodic magnetization configurations (see Supplementary Note 2 for details). A nonzero Bott index $B$ without an external magnetic field indicates that the electronic fluid responds to the internal fields generated by nontrivial spin-textures. In the absence of inhomogeneities, Bott index $B$ equals the Chern number, a well-known invariant for topological insulators, thereby linking real-space and momentum-space topologies.

**Distinct topological phases after field cooling.**

Distinct topological phases in an FGT flake are accessed by thermal cycling across the $T_C$ to measurement temperatures of 194 and 130 K (Fig. 2a). Zero-field cooling ($H_{fc}$ = 0 mT) yields labyrinthine domains; at $H_{fc}$ = -8 mT a mixed labyrinth-skyrmion phase is observed; and at $H_{fc}$ = -16 mT isolated skyrmions appear. Each configuration is consistently observed at both temperatures. Theoretical simulations reproduced these experimental textures with high fidelity, validating our atomistic description for the diverse magnetic phases in FGT.

**Skyrmion nucleation and annihilation during the temperature loop.**

To examine temperature-driven behavior, we focus on $H_{fc}$ = -16 mT and track skyrmion evolution across $T_C$ (Fig. 2b). The skyrmion density decreases with cooling, and the state evolves symmetrically during cooling and warming, indicating a robust formation mechanism resilient to thermal perturbations. The simulations corroborate these observations, showing thermally induced elongation of skyrmion bubbles with increasing temperature. Notably, intensity smearing appears at 194 K and is absent at 130 K in the experimental images (both Figs. 2a and b), which is interpreted as resulting from fluctuating domain states at elevated temperatures, as analyzed with coherent correlation imaging (CCI)[46], as described later. Figure 2c shows the temperature dependence of the skyrmion diameter $D_s$ and the skyrmion density $\chi_s$, derived from real-space configurations and Eqs. 8-9, respectively (detail in Method and Supplementary note 1). The decrease in diameter $D_s$ with increasing temperature accompanies a rise in density $\chi_s$, consistent with the experimental results.

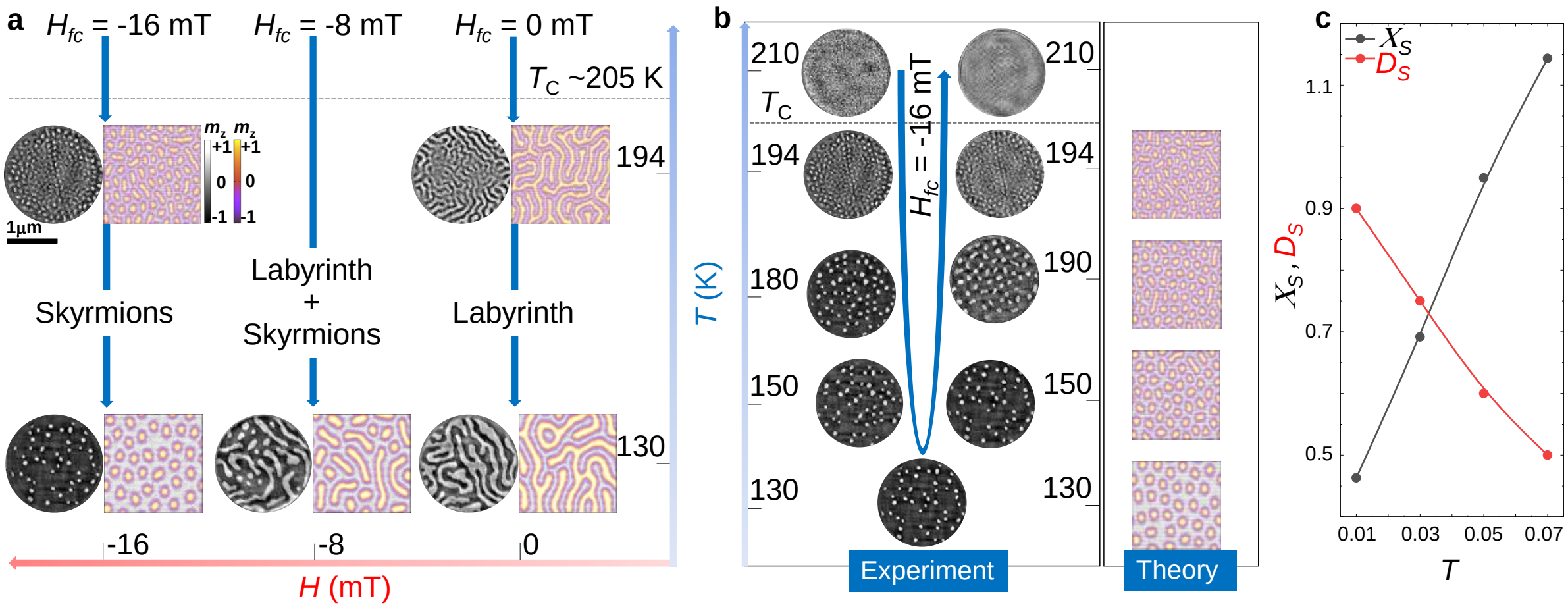

**Figure 2| Distinct topological phases and thermally symmetric evolution of skyrmions. a,** Spin-textures in FGT after field cooling through $T_C$ at different applied fields and temperatures. Theoretical simulations were conducted for fixed $T = 0.01$, $\lambda = 0.2$, and $A_u = 0.1$, with varying field $H_z$ = 0, -0.012, -0.025, corresponding to labyrinth, mixed, and skyrmion phases at 130 K.

Simulations for $T = 0.07$, $\lambda = 0.3$, and $A_u = 0.12$, with $H_z = 0$ and -0.05, correspond to labyrinth and skyrmion phases at 194 K. **b,** Temperature evolution of skyrmions during both cooling and warming across $T_C$ under $H_{fc} = -16$ mT. The simulation parameters (bottom to top) are: $\lambda = 0.2$, $T = 0.01$, $H_z = -0.025$, $A_u = 0.10$ (130 K); $\lambda = 0.233$, $T = 0.03$, $H_z = -0.033$, $A_u = 0.1067$ (150 K); $\lambda = 0.267$, $T = 0.05$, $H_z = -0.042$, $A_u = 0.1133$ (190 K); and $\lambda = 0.3$, $T = 0.07$, $H_z = -0.05$, $A_u = 0.12$ (194 K). **c,** Temperature dependence of skyrmion diameter ($D_s$) and skyrmion density ($\chi_s$), in units of lattice spacing and count per unit-cell area, respectively. The $\chi_s$ and $D_S$ are multiplied by 25 and 0.05, respectively, to bring them to the same scale.

**Distinct evolution of topologically different phases.**

Next, we examined the field-dependent evolution of these topological phases at 130 K after cooling through $T_C$ with $H_{fc} = 0$, -8, and -16 mT (Fig. 3). For $H_{fc} = 0$ mT, the system begins with a labyrinthine domain structure that gradually evolves into a single-domain ferromagnetic state as the field increases (Fig. 3a). Repeated measurements show spatial patterns shifting to different positions within the same field of view while retaining their overall labyrinthine character (Supplementary Fig. 3), indicating that phase evolution is only weakly affected by pinning. The evolution of spin-textures with applied field is quantified by skyrmion density $\chi_s$, plotted as a function of the field in Fig. 3b; the field does not induce a substantial increase in $\chi_s$, which remains consistently low.

For $H_{fc} = -8$ mT, intermediate fields (e.g., 40, 56, 72 mT) reveal skyrmion bag configurations in both experiment and simulation (Fig. 3c)[8,42,67,68]. Similar features appear for $H_{fc} = -4$ mT (Supplementary Fig. 4), where the initial state again comprises a mixed labyrinth-skyrmion phase. These results indicate that skyrmion bag states emerge generically during field evolution from mixed labyrinth-skyrmion topological states in FGT. The density $\chi_s$ exhibits a gradual increase with field, followed by a sharp drop at higher fields (Fig. 3d).

For $H_{fc} = -16$ mT, the system starts with isolated skyrmions that progressively expand into larger magnetic islands as the field increases (Fig. 3e). A substantial skyrmion population exists even at zero field and remains nearly constant over a range of field values before gradually dropping to zero at higher fields, marking the disappearance of skyrmions (Fig. 3f). Across all three conditions, the application of sufficiently large fields drives the system into a uniform ferromagnetic state with vanishing skyrmion density $\chi_s$.

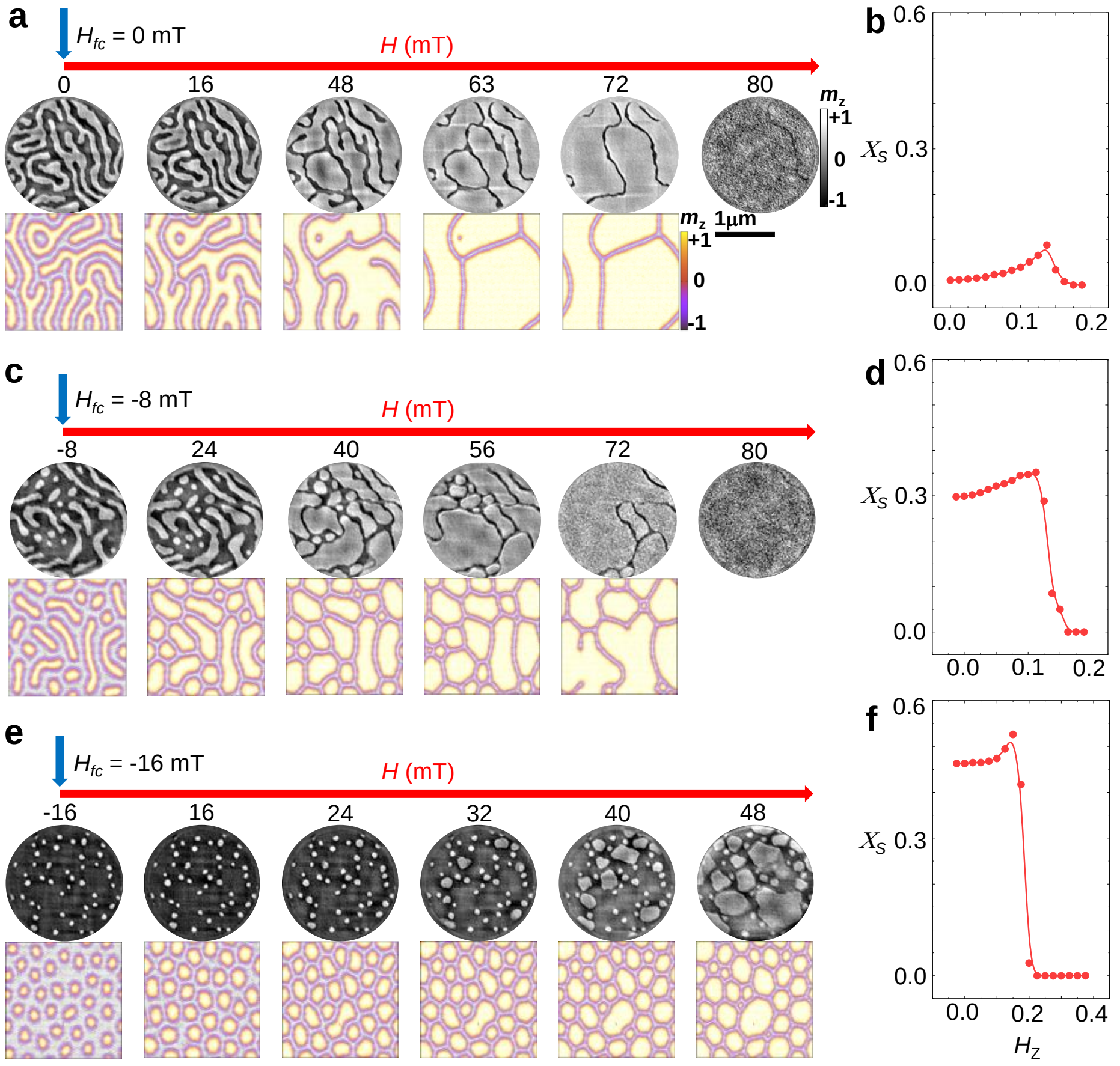

**Figure 3| Field-dependent evolution of topological phases in FGT. a, c, e,** Spin-textures showing the evolution of magnetic domains in the FGT thin flake at 130 K after field cooling under different magnetic fields: **a,** $H_{fc}$ = 0 mT; **c,** $H_{fc}$ = -8 mT; and **e,** $H_{fc}$ = -16 mT. Theoretical simulations were conducted by varying the external magnetic field $H_Z$ while keeping $T = 0.01$, $\lambda = 0.2$, and $A_u = 0.1$ constant. **b, d, f,** corresponding theoretical variation of skyrmion density $\chi_s$, in the unit of count per unit-cell area, with $H_Z$ for each cooling condition shown in panels **a**, **c**, and **e**, respectively. The multiplication factor for the $\chi_s$ is 25.

**Topological transitions and protection: near $T_C$ vs. fer below $T_C$.**

We examine the behavior and stability of topological phase transitions by imaging spin-textures during magnetic-field cycling across temperatures. Figure 4a shows the evolution at 130 K (well below $T_C$) after field cooling through $T_C$ at $H_{fc}$ = -16 mT. As the applied field increases from -16 to 48 mT, isolated skyrmions expand into larger magnetic puddle-like structures; some persist, while others merge to form irregular islands. Reversing the field from 48 to -63 mT contracts these islands, and the skyrmions are progressively annihilated, yielding a uniform magnetic domain state. Theoretical simulations (Fig. 4b) reproduce this evolution with high fidelity.

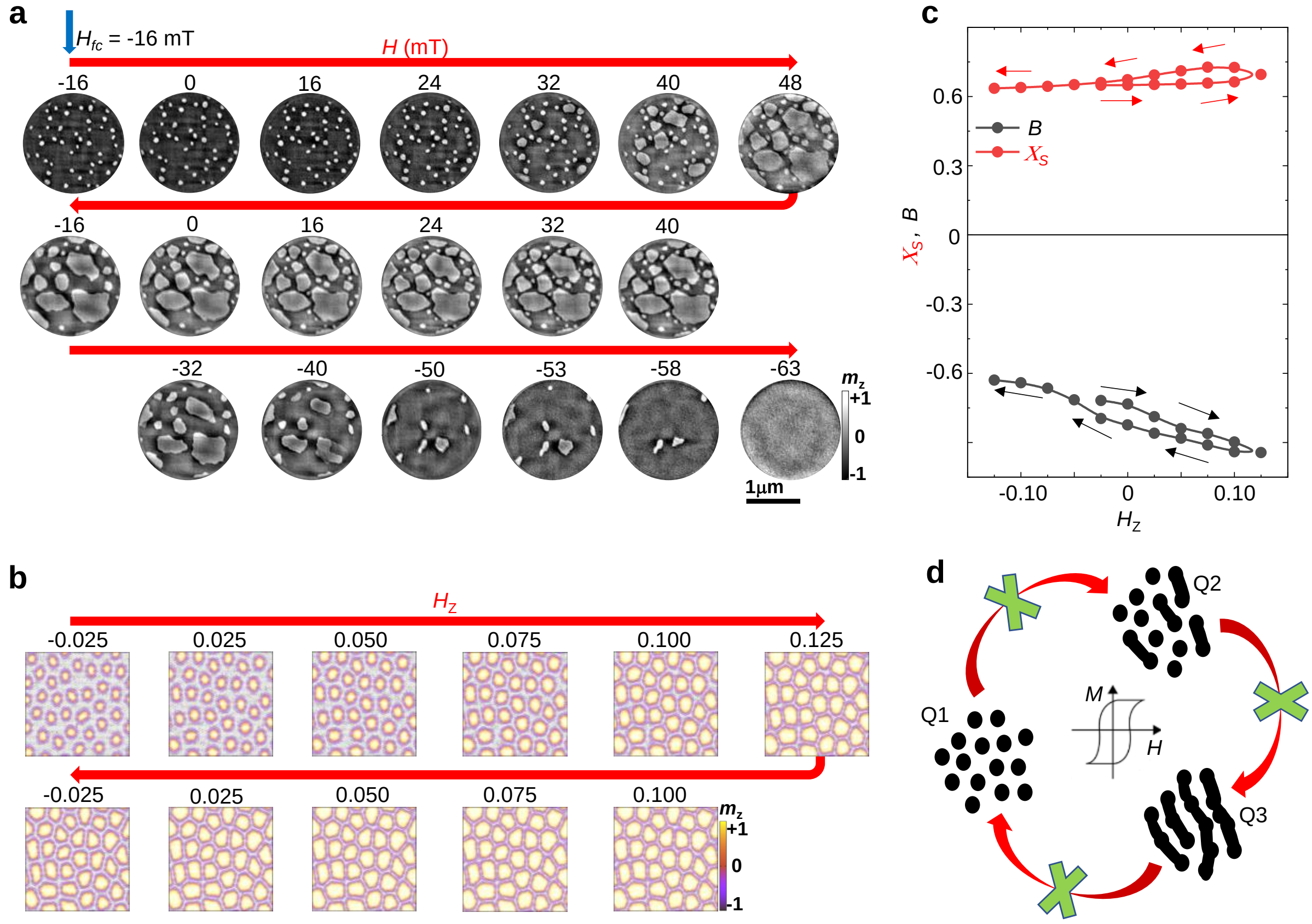

**Fig. 4| Absence of topological transitions and protective topological state at far below $T_C$ (130 K). a,** Experimental spin-texture in an FGT thin flake during magnetic-field cycling at 130 K, after field cooling through $T_C$ at $H_{fc}$ = -16 mT. **b,** Theoretical real-space spin-textures for a system of size $N$ =120², after field cooling at $H_Z$ = -0.025 ($T$ = 0.01, $\lambda$ = 0.2, $A_u$ = 0.10). **c,** Skyrmion density $\chi_s$ and Bott index $B$ as a function of $H_Z$. $\chi_s$ is in the unit of count per unit-cell area. $\chi_s$ is scaled by 25 for comparability. **d,** Schematic of topological non-transitions and protected topological states under field cycling. $Q_1$, $Q_2$, and $Q_3$ denote distinct spin-configurations with different topological charges; $M$ and $H$ denote magnetization and magnetic field, respectively.

Topologically nontrivial textures appear in both insulators and metals. In insulators, such textures generally do not influence electronic behavior, whereas in metals their impact depends on the coupling strength between magnetic and electronic degrees of freedom. Weak coupling yields little effect, while strong coupling can significantly alter the electronic response, motivating spintronic applications where magnetic fields control electronic properties. Consequently, skyrmion density alone is insufficient to distinguish different skyrmion-hosting materials. To quantify the electronic fluid state, we compute the Bott index $B$ and the skyrmion density $\chi_s$. The Bott index, suited for magnetically non-periodic or disordered systems, is equivalent to the Chern number in insulating systems and is sensitive to the electronic response to inhomogeneous magnetic textures in metals[60-62] (see Supplementary Note 2 for details). Figure 4c shows that the Bott index $B$ and the skyrmion density $\chi_s$ exhibit a consistent overall topological evolution during field cycling, although their detailed field dependence is not identical. Importantly, neither quantity crosses zero, indicating the absence of a topological transition and the persistence of protected topological states. A schematic illustrating this behavior is shown in Fig. 4d.

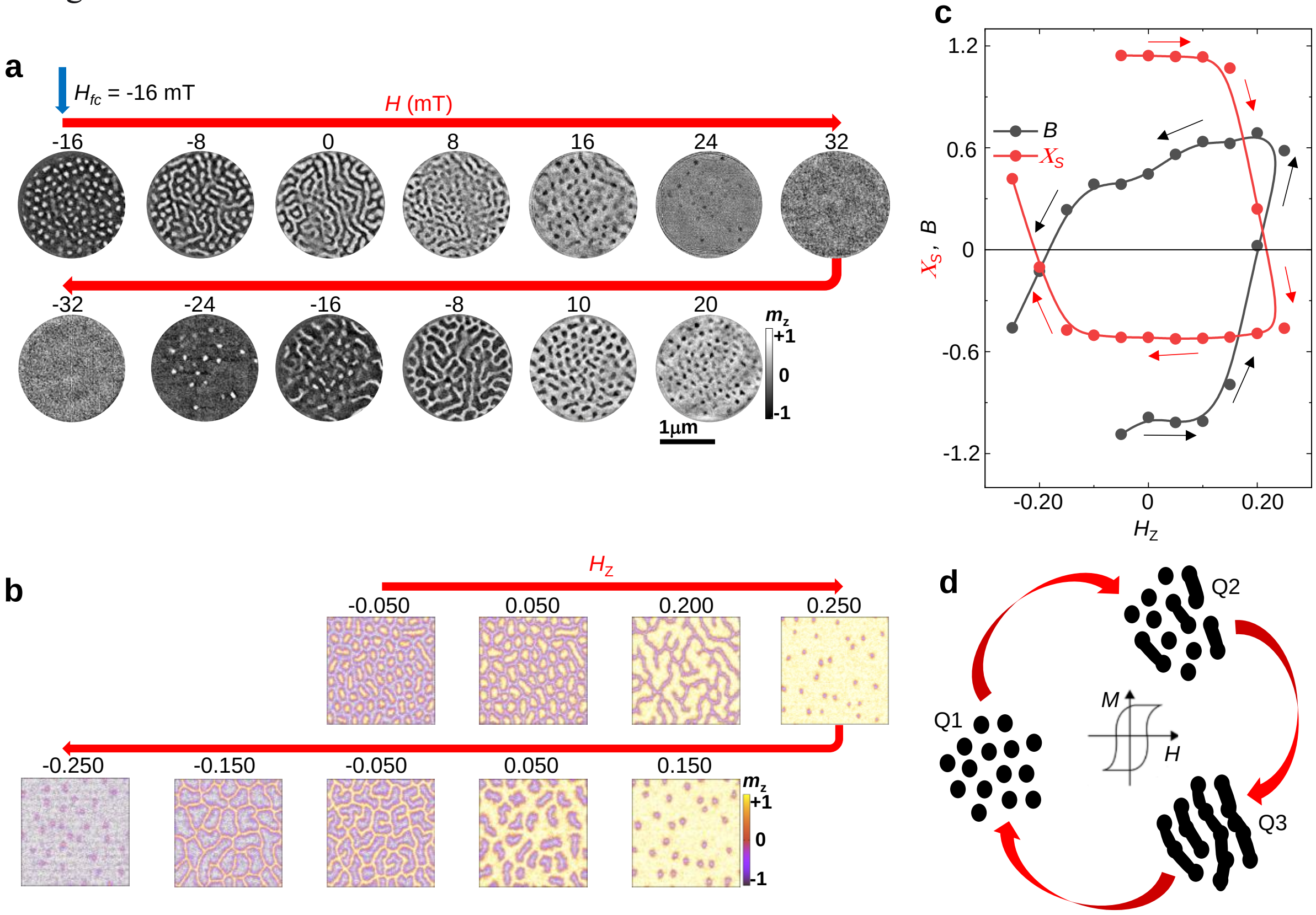

**Figure 5| Topological transitions and unprotected topological states near $T_C$ (194 K). a,** Experimental spin-texture in the FGT thin flake during magnetic-field cycling at 194 K after field cooling through $T_C$ at $H_{fc}$ = -16 mT. **b,** Theoretical real-space spin-textures for a $N = 120^2$ system, after field cooling at $H_Z$ = -0.050 ($T$ = 0.07, $\lambda$ = 0.3, $A_u$ = 0.12). **c,** Variation of skyrmion density $\chi_s$ and Bott index $B$ as a function of $H_Z$. $\chi_s$ is scaled by 25 for comparability. **d,** Schematic illustrating topological transitions and unprotected topological states under field cycling.

In contrast, field cycling at 194 K (near $T_C$) reveals markedly different behavior (Fig. 5a). Starting from $H_{fc}$ = -16 mT, skyrmions merge into mixed skyrmion-labyrinth textures under applied fields (e.g., -8 mT), which subsequently evolve into uniform labyrinth domains at 0 mT. Reversing the field, these textures re-emerge in reverse order, demonstrating a reversible cyclic transformation among pure skyrmions, labyrinth domains, and mixed phases. The simulations (Fig. 5b) closely reproduce this cyclic behavior. As shown in Fig. 5c, both $\chi_s$ and $B$ exhibit a clear zero-line crossing: $\chi_s$ transitions from positive to negative values around $H_Z \approx$ 0.2, accompanied by a corresponding sign reversal in $B$. This behaviour indicates a thermally driven topological transition and highlights the existence of topologically unprotected states near $T_C$. Figure 5d schematically summarizes these transitions and unprotected states. Black dots denote skyrmions. Field variation in one direction promotes the merging (coalescence) of skyrmions into a mixed labyrinth-skyrmion state, followed by cycling that yields a pure labyrinth texture. Sweeping the field in the opposite direction yields cyclic reappearance of the spin-textures, as observed in Figs. 5a and b.

**Discussion and Outlook**

We find that labyrinthine domains, isolated skyrmions, and mixed labyrinth-skyrmion phases can be deterministically stabilized in FGT thin flake by specific field-cooling protocols, both near $T_C$ and well below $T_C$ (Fig. 2a). Near $T_C$, these spin-textures exhibit dynamic interplay under magnetic-field cycling, yielding reversible transitions between distinct topological configurations and the emergence of thermally unprotected states (Fig. 5). A similar field-induced switching is observed after $H_{fc}$ = 0 mT through $T_C$, starting with labyrinth phase (Supplementary Fig. 5), indicating universality of topological transitions and the presence of unprotected topological states near $T_C$, independent of the initial magnetic-texture. In contrast, far below $T_C$, tunability is lost: the system responds asymmetrically to field cycling and stabilizes into fixed, non-interconvertible textures (Fig. 4).

Crucially, we identify topological transitions using topology-sensitive metrics, skyrmion density, and the Bott index, beyond visual inspection of real-space textures. In metallic FGT, these transitions are reflected in the evolution of the topological contribution to the Hall conductivity during magnetic-field cycling. Our simulations reveal that the qualitative difference between field cycling well below $T_C$ and near $T_C$ arises from a rugged energy landscape containing high-energy intermediate configurations that must be traversed to switch between topological phases; such intermediates are suppressed at low temperatures. Near $T_C$, enhanced entropy in the free energy enables access to these intermediates, allowing barrier crossing and stabilization of distinct low-energy topological states. Proximity to $T_C$ thus enables the topological switching we observe.

The temperature dependence of skyrmions can be understood in terms of an effective renormalization of the electronic bandwidth (or hopping amplitude) within a double-exchange framework, arising from the coupling between itinerant electrons and local moments. In FGT, itinerant electrons are strongly coupled to local moments[53,69], so the electronic bandwidth, and hence the effective ferromagnetic exchange, decreases monotonically across the ferromagnetic-paramagnetic transition. We implement this by introducing a temperature dependence in the hopping parameter $t$. In our model, this is captured by introducing a temperature dependence of the effective hopping parameter $t$, which reflects the renormalization of magnetic energy scales rather than a direct modification of the bare band structure. Such renormalization of effective kinetic energy scales with temperature is a well-established feature of double-exchange systems. This separation of energy scales between the electronic bandwidth and the magnetic ordering temperature is consistent with expectations for double-exchange systems[69]. As a result, Rashba SOC, magnetic anisotropy, and the effective field strength gain relative importance, yielding smaller skyrmions and higher skyrmion densities at higher temperatures (Fig. 2c). The interpretation aligns with both experimental observations and simulations (Fig. 2).

We identify the Rashba SOC-to-bandwidth ratio as a key control parameter in the formalism. In FGT, anisotropic DMI-like and dipolar-like interactions, typically introduced as independent inputs in micromagnetic simulations, emerge naturally from this ratio (see Supplementary Note 2 for details). Because bandwidth in metallic systems can be tuned by various mechanisms (e.g., gating, carrier density, or strain), our findings point to practical routes for engineering magnetic-textures.

Phenomenologically, the diverse skyrmion states in FGT have been attributed to the interplay between dipolar interactions and magnetocrystalline anisotropy[7,38]. Conventional micromagnetic simulations based on these ingredients neglect a quantum-mechanical description of itinerant electrons and therefore cannot capture the electronic response to external perturbations such as electric and magnetic fields. Moreover, the coarse-grained magnetization description in these models becomes unreliable for skyrmion radii below ~100 nm. By contrast, our atomic-scale lattice Hamiltonian explicitly accounts for FGT's metallic character and its impact on magnetic interactions, providing a natural framework to describe these effects.

Near $T_C$, we observe smeared intensity in the experimental spin-textures (Figs. 2a-b, Fig. 5a, and Supplementary Fig. 5a), appearing as grey regions that reflect fluctuations of domain configurations during data acquisition at elevated temperatures. This effect is absent at 130 K, indicating that domain configurations remain static there. To capture this dynamic behavior, we performed CCI[46] analysis on the -8 mT data at 194 K in Fig. 5a as an example (as shown in Supplementary Fig. 6; detail in Methods), which reveals temporal rearrangements of the magnetic-textures throughout the measurement. These results highlight the dynamic character of textures near the phase transition and point to promising opportunities for future correlation spectroscopy studies of domain stability in FGT.

Our integrated experimental-theoretical framework combines drift-free, high-resolution X-ray magnetic imaging with Monte Carlo simulations based on a Rashba-type tight-binding model to probe topological transitions and their microscopic origins in skyrmion-hosting metallic magnets, notably FGT. Nonzero Bott indices, a topological invariant for electronic degrees of freedom in non-periodic systems, confirm a dual topological character: the real-space topology of spin-textures is imprinted on the electronic band structure beyond the scope of classical descriptions. Excellent agreement between simulation and experiment across a range of temperatures and magnetic fields in sequential variation supports a mechanism in which electronic bandwidth and SOC govern texture formation. This insight suggests practical routes[70,71] to tune skyrmion size, density, and stability via bandwidth-control strategies (e.g., isoelectronic doping or pressure). Moreover, tailored field-cooling protocols reproducibly generate specific magnetic textures together with their electronic responses. Taken together, these results establish a versatile foundation for topology-driven logic and non-volatile memory technologies.

## Methods

**A. Single crystal preparation:** Single crystals of FGT were synthesized by chemical vapor transport (CVT) using $TeCl_4$ as the transport agent (in contrast to our previous reports that used $I_2$)[72-74]. High-purity precursors Fe (Alfa Aesar, 99.99%), Ge (Merck, 99.999%), and Te (G-Materials, 99.999%) were weighed in a 3.1:1:2 molar ratio, ground in an argon-filled glove box, mixed with $TeCl_4$, and sealed in evacuated fused silica ampoules (~$10^{-5}$ mbar). Each ampoule (~20 cm long) contained ~3.0 g of the mixture and ~1 mg/cm$^3$ of $TeCl_4$. The ampoules were placed in a two-zone furnace for 12 days, with the hot end at ~745 °C and the cold end at ~695 °C. Shiny, hexagonal crystals were harvested from the cold end and are readily cleavable with Scotch™ tape to expose fresh surfaces for processing. Elemental composition was confirmed by energy-dispersive X-ray spectroscopy (EDS) on a COXEM EM-30plus scanning electron microscope (SEM) with an Oxford Silicon Drift Detector (SDD) and AZtecLiveLite software, yielding Fe:Ge:Te ≈ 2.96:1.00:2.13 (Supplementary Fig. 1c), indicating near-stoichiometric composition.

**B. Magnetometry:** Magnetic properties of FGT single crystals were characterized using a superconducting quantum interference device (SQUID) magnetometer with applied fields up to 7 T. The temperature dependence of the magnetic volume susceptibility $\chi_V(T)$ was measured under a field-cooled protocol at $H = 0.1$ T, and the isothermal magnetization $M(H)$ was recorded at 2 K. These measurements are shown in Supplementary Fig. 1a (susceptibility) and Supplementary Fig. 1b ($M$-$H$), respectively. For both protocols, the magnetic field was applied along the *ab*-plane ($H \parallel ab$) and along the *c*-axis ($H \parallel c$), enabling assessment of anisotropic magnetic behavior.

**C. Holography mask preparation:** Prior to sample transfer, a holography mask was prepared on the backside of silicon nitride ($Si_3N_4$) holders (150 nm thick). The mask features a circular aperture (1.9 µm diameter) milled by focused-ion-beam (FIB) into a [Cr(5 nm)/Au(50 nm)]×13 multilayer, which remains opaque to soft X-rays except at the aperture. The aperture stops at the $Si_3N_4$ membrane and defines the fixed microscopic field of view. After flake transfer, five reference pinholes (diameter < 100 nm) were milled by FIB through the sample to provide reference beams for holographic imaging (Supplementary Fig. 2).

**D. Thin flake preparation:** FGT thin flake was mechanically exfoliated from bulk single crystals using Scotch™ tape and initially transferred onto a polydimethylsiloxane (PDMS) film on a glass slide. Flake with a thickness of approximately 150 nm was identified by optical contrast and subsequently transferred onto the pre-fabricated holography mask using a custom-built micro-aligner. To prevent oxidation, the flake was immediately encapsulated with about 20 nm of hexagonal boron nitride (*h*BN) via the same transfer route. The total time from exfoliation to encapsulation was kept under 10 minutes to minimize air exposure.

**E. Fourier Transform Holography (FTH); X-ray Imaging:** Coherent imaging experiments were conducted at the MAX-P04 endstation, beamline P04, PETRA III (Hamburg, Germany). The circularly polarized X-ray beam was tuned to the Fe $L_3$-edge ($\lambda = 1.75$ nm; E ≈ 708 eV) to exploit X-ray magnetic circular dichroism (XMCD) as the magnetic-contrast mechanism. Scattered radiation from the sample was recorded with a back-illuminated CCD camera (PI SOPHIA XO; 2048 × 2048 pixels; 15 µm pixel size) positioned 205 mm from the sample. Real-space images of the out-of-plane magnetization were reconstructed from the difference between two measurements made with opposite X-ray helicities. Reconstruction employed FTH with numerical phase retrieval.

**F. Coherent correlation imaging (CCI).** For each helicity, 70 frames were acquired with a CCD detector, with a one-minute pause during helicity reversal. The CCD exposure time was 0.7 s and the readout time 1.0 s, yielding a temporal resolution of 1.7 s per frame; a complete image at a given field and temperature required ~300 s. The CCI[46] analysis grouped the frames into four-time segments: frames 0-44 (0-74.8 s), 45-70 (74.9-119 s), 71-119 (179-264.3 s), and 120-140 (264.4-298 s), assigned as segments 1-4 in Supplementary Fig. 6. This segmentation captures temporal rearrangements of the magnetic textures during the measurement.

**G. Effective spin model and Monte Carlo simulations:** The starting point for the theoretical investigations is an electronic lattice Hamiltonian with Hund's rule coupling, Rashba spin-orbit coupling, and uniaxial anisotropy. The model is given by:

$$\begin{aligned} H &= -t\sum_{i,\gamma,\sigma}(c^{\dagger}_{i,\sigma}c_{i+\gamma,\sigma}+\mathrm{H.c.})-\sum_{i}[h_z S^z_i + A_u(S^z_i)^2] \\ &\quad - J_{\mathrm{H}}\sum_{i,\gamma,\sigma\sigma'} c^{\dagger}_{i\sigma}[\boldsymbol{\tau}\cdot(\mathbf{S}_i + \mathrm{i}\frac{\lambda}{J_{\mathrm{H}}}(\hat{\boldsymbol{\gamma}}\times\hat{\mathbf{z}}))]_{\sigma\sigma'}c_{j\sigma'}. \end{aligned} \tag{1}$$

Here, $c_{i\sigma}$ and $c^{\dagger}_{i\sigma}$ are respectively the electronic annihilation and creation operators, $t$ is the nearest-neighbor hopping amplitude, and $J_{\mathrm{H}}$ $(\lambda)$ denotes the strength of Hund's (Rashba) coupling, $\boldsymbol{\tau}$ is a vector operator with the three Pauli matrices as components. The localized spins, $\mathbf{S}_i$ , are treated classically. $A_u$ is the easy-axis uniaxial anisotropy parameter and $h_z$ is the strength of the external magnetic field. Assuming the lattice constant to be unity, $\hat{\boldsymbol{\gamma}}$ denotes the unit vector along the primitive vectors of the triangular Bravais lattice. All energy scales of the Hamiltonian are measured in units of $t$. For the case of Rashba coupling relevant for thin films, the effective spin Hamiltonian is given by[75]:

$$\begin{aligned} H_{\mathrm{eff}} &= -\textstyle\sum_{\langle ij\rangle,\gamma} D^{\gamma}_{ij} f^{\gamma}_{ij} - \sum_i [h_z S^z_i + A_u (S^z_i)^2] \\ \sqrt{2} f^{\gamma}_{ij} &= \big[t^2(1+\mathbf{S}_i\cdot\mathbf{S}_j) + 2t\lambda\hat{\boldsymbol{\gamma}}'\cdot(\mathbf{S}_i\times\mathbf{S}_j) \\ &\quad +\lambda^2(1-\mathbf{S}_i\cdot\mathbf{S}_j + 2(\hat{\boldsymbol{\gamma}}'\cdot\mathbf{S}_i)(\hat{\boldsymbol{\gamma}}'\cdot\mathbf{S}_j))\big]^{1/2}, \\ D^{\gamma}_{ij} &= \langle [e^{\mathrm{i}h^{\gamma}_{ij}} d^{\dagger}_i d_j + \mathrm{H}.c.]\rangle_{gs} \end{aligned} \tag{2}$$

In the above, $f^{\gamma}_{ij}(h^{\gamma}_{ij})$ is the modulus (argument) of the complex number $g^{\gamma}_{ij}$, and $\langle\hat{O}\rangle_{gs}$ denotes expectation values of the operator $\hat{O}$ in the ground state. We perform Monte Carlo simulations on the effective spin Hamiltonian via the standard Metropolis algorithm[48-53].

The presence of the skyrmion Hall effect in the vdW magnets indicates that the strong-coupling limit is more relevant here as compared to the weak-coupling RKKY limit[48-51]. So, in the strong-coupling limit, Eq. (1) leads to the Rashba double-exchange (RDE) Hamiltonian on a triangular lattice[52],

$$H_{\mathrm{RDE}} = \textstyle\sum_{\langle ij\rangle,\gamma}[g^{\gamma}_{ij} d^{\dagger}_i d_j + \mathrm{H.c.}] - \sum_i [h_z S^z_i + A_u (S^z_i)^2] \tag{3}$$

where, $d_i(d^{\dagger}_i)$ annihilates (creates) an electron at the site i with spin parallel to the localized spin. Site $j = i + \boldsymbol{\gamma}$ is the nn of sites i along one of the three symmetry directions on the triangular lattice. The projected hopping $g_{ij}^{\gamma}$ depend on the orientations of the local moments $\mathbf{S}_i$ and $\mathbf{S}_j$. The tight-binding, $t_{ij}^{\gamma}$ and Rashba, $\lambda_{ij}^{\gamma}$, contributions to $g_{ij}^{\gamma} = t_{ij}^{\gamma} + \lambda_{ij}^{\gamma}$ are given by[52],

$$t_{ij}^{\gamma} = -t\left[\cos(\tfrac{\theta_i}{2})\cos(\tfrac{\theta_j}{2}) + \sin(\tfrac{\theta_i}{2})\sin(\tfrac{\theta_j}{2})e^{-\mathrm{i}(\phi_i-\phi_j)}\right],$$

$$\lambda_{ij}^{\mathbf{a}_1} = \lambda_{ij}^{x}, \qquad \lambda_{ij}^{\mathbf{a}_{2/3}} = \pm\tfrac{1}{2}\lambda_{ij}^{x} + \tfrac{\sqrt{3}}{2}\lambda_{ij}^{y},$$

$$\lambda_{ij}^{x} = \lambda\left[\sin(\tfrac{\theta_i}{2})\cos(\tfrac{\theta_j}{2})e^{-\mathrm{i}\phi_i} - \cos(\tfrac{\theta_i}{2})\sin(\tfrac{\theta_j}{2})e^{\mathrm{i}\phi_j}\right],$$

$$\lambda_{ij}^{y} = \mathrm{i}\lambda\left[\sin(\tfrac{\theta_i}{2})\cos(\tfrac{\theta_j}{2})e^{-\mathrm{i}\phi_i} + \cos(\tfrac{\theta_i}{2})\sin(\tfrac{\theta_j}{2})e^{\mathrm{i}\phi_j}\right] \tag{4}$$

where $\theta_i$ $(\phi_i)$ is the polar (azimuthal) angle for the localized moment $\mathbf{S}_i$.

The Hamiltonian Eq. (3) describes a modified tight-binding model where the hopping integrals are dependent on the configuration of classical spins. Therefore, the energy of the system depends on the classical spin configurations. This dependence can be formally written as an effective spin Hamiltonian by following a procedure well known for double-exchange models[52,53], and is given by Eq. (4).

Assuming constant coupling parameters is a good approximation for studying the ground-state phases of $H_{\text{eff}}$[52], therefore, we set $D_{ij}^{\gamma} \equiv D_0 = 1/\sqrt{2}$ to study $H_{\text{eff}}$. We perform Monte Carlo simulations on Hamiltonian Eq. (2) via the standard Markov chain Monte Carlo using the Metropolis algorithm. To follow closely the experimental results, we use multiple protocols for varying temperatures and magnetic fields during simulations. Typically, we begin the simulations at high temperature and use $10^5$ Monte Carlo steps for equilibration and an equal number for averaging at each parameter point. Most importantly, the Monte Carlo approach allows access to full spatial details that can be used not only to identify unusual magnetization textures but also to compare with spatially resolved experimental data, as we discuss later. Therefore, our analysis focuses on the typical spin configurations and their evolution with magnetic field cycles and with temperature.

The theoretical model used here is a minimal effective tight-binding description intended to capture the essential low-energy ingredients relevant for spin-texture formation in FGT, namely itinerant electrons, SOC, Hund's coupling, and magnetic anisotropy. It is not meant to reproduce the full multiband density functional theory (DFT) electronic structure quantitatively. Rather, it is motivated by previous DFT and angle-resolved photoemission spectroscopy (ARPES) studies showing that FGT is a correlated Hund's metal with strong SOC and unconventional band topology, including nodal-line features[40,72,76]. These ingredients are incorporated here at an effective level through a one-band tight-binding Hamiltonian with Rashba-type SOC.

**H. Bott index and skyrmion density calculations for topological characterization:** The Bott index is a way to measure how well a system can or can't form localized Wannier orbitals from its occupied states[60]. Here, we calculate the Bott index to confirm, in a precise mathematical way[61,62], the topological nature of the electronic band structure when magnetic skyrmion states are present. The first step is to construct a projection operator out of all the occupied states:

$$\mathrm{P}=\sum_{k=1}^{Nel} |\psi_k\rangle \langle\psi_k| \tag{5}$$

where, $|\psi_k\rangle$ is the occupied eigenstate associated with the k-th eigenvalue $\mathrm{E}_k$, and $N_{el}$ is the number of electrons in the system. The position coordinates $(x_i, y_i)$ of any lattice site i can be mapped into the spherical coordinates $(\Theta_i, \Phi_i)$ on a torus where $0 \leq \Theta_i < 2\pi$ and $0 \leq \Phi_i < 2\pi$. The next step involves defining the projected position operators:

$$U = Pe^{i\Theta}P,$$

$$V = Pe^{i\Phi}P \tag{6}$$

where, $\Theta$ and $\Phi$ are the diagonal matrices with $\Theta_i$ and $\Phi_i$ as diagonal elements respectively. The Bott index is given by

$$B = \frac{1}{2\pi}\,\mathrm{Im}\left\{\mathrm{tr}\left[\log\left(VUV^{\dagger}U^{\dagger}\right)\right]\right\} \tag{7}$$

To keep the calculations stable, we use singular value decomposition (SVD) on the projected position operators U and V, as described by Huang and Liu[63,64]. For this, we make use of the LAPACK library: the 'CGEES' subroutine handles the diagonalization of complex, non-symmetric matrices, and the 'CGESVD' subroutine takes care of the SVD.

We have also calculated the average skyrmion density, which can be given by[65]:

$$\chi_s = \frac{1}{4\pi N}\sum_i \left(A_i^{(12)}\ \mathrm{sgn}\left[\mathcal{L}_i^{(12)}\right] + A_i^{(45)}\ \mathrm{sgn}\left[\mathcal{L}_i^{(45)}\right]\right) \tag{8}$$

$$\mathcal{L} = \frac{1}{8\pi}\left\langle \sum_i \left(\mathcal{L}_i^{(12)} + \mathcal{L}_i^{(45)}\right)\right\rangle \tag{9}$$

where, $A_i^{(ab)} = \|(\mathbf{S}_i^a - \mathbf{S}_i) \times (\mathbf{S}_i^b - \mathbf{S}_i)\|/2$ is the local area of the surface spanned by three spins on every elementary triangular plaquette $\mathbf{r}_i, \mathbf{r}_a, \mathbf{r}_b$. Here $\mathcal{L}_i^{(ab)} = \mathbf{S}_i \cdot (\mathbf{S}_i^a \times \mathbf{S}_i^b)$ is the so-called local chirality, and the numbers 1-6 are the nearest neighbour sites of a site *i* taken in an anti-clockwise manner, which are involved in the calculation of $\chi_s$.

Here, the Bott index is a theoretically computed quantity obtained from the simulated electronic states corresponding to the spin-textures. Direct experimental determination of its sign would require complementary Hall transport measurements[66] and is beyond the scope of the present study.

**Acknowledgements.** The measurements presented in this work were carried out at PETRA III (DESY, Hamburg, Germany). DESY, a member of the Helmholtz Association (HGF), is gratefully acknowledged for providing access to the PETRA III facilities at beamline P04. We thank Frank Scholz and Arne Meyer for their expert technical assistance. S.R.D. and S.K. recognize the use of the High-Performan ce Computing facility at IISER Mohali; S.R.D. acknowledges financial support from IISER Mohali through the institute fellowship. T.A.B. acknowledges funding from the Swiss Nanoscience Institute (SNI) and the European Regional Development Fund (ERDF). K.W. and T.T. acknowledge support from the JSPS KAKENHI (Grant Numbers 21H05233 and 23H02052), the CREST (JPMJCR24A5), JST, and World Premier International Research Center Initiative (WPI), MEXT, Japan.

**Author contributions.** S.C., S.K., A.-A.H., and M.H. conceived the project and directed the work. S.P. and A.-A.H. grew the single crystals and performed EDX and magnetometry. C.H.S. prepared the thin flake. M.S. prepared the holography mask and carried out FIB milling. S.C., M.S., C.K., L.-M.K., T.A.B., J.F., and B.P. performed the X-ray imaging beamtime. S.R.D. performed the Monte Carlo simulations under the supervision of S.K. S.C., and S.R.D. analyzed the data and prepared the figures. S.C., S.R.D., S.K., S.D., B.P., A.-A.H., and M.H. interpreted the data. S.C., S.R.D., S.D., and S.K. wrote the original manuscript with the assistance of S.D.G., B.P., A.-A.H., and M.H. All authors discussed the results and commented on the manuscript.

**Competing interests.** There are no competing interests to declare.

**Data availability.** The data that support the findings of this study are available from the corresponding authors upon reasonable request.

**References**

1. D. R. Klein *et al.*, *Science* **2018**, *360*, 1218–1222.
2. T. Song *et al.*, *Science* **2018**, *360*, 1214–1218.
3. P. Pal *et al.*, *arXiv* **2024**, arXiv:2410.22447.
4. W. Jiang *et al.*, *Science* **2015**, *349*, 283–286.
5. S. Woo *et al.*, *Nat. Mater.* **2016**, *15*, 501–506.
6. F. Büttner *et al.*, *Nat. Mater.* **2021**, *20*, 30–37.
7. M. T. Birch *et al.*, *Nat. Commun.* **2022**, *13*, 3035.
8. L. Powalla *et al.*, *Adv. Mater.* **2023**, *35*, 2208930.
9. Y. Zhou, S. Li, X. Liang, Y. Zhou, *Adv. Mater.* **2025**, *37*, 2312935.
10. S. S. P. Parkin, M. Hayashi, L. Thomas, *Science* **2008**, *320*, 190–194.
11. S. Manipatruni *et al.*, *Nature* **2019**, *565*, 35–42.
12. J. Pei *et al.*, *Nature* **2019**, *572*, 106–111.
13. Y. Deng *et al.*, *Nature* **2018**, *563*, 94–99.
14. D. Z. Plummer *et al.*, *arXiv* **2024**, arXiv:2503.17376.
15. S. Mühlbauer *et al.*, *Science* **2009**, *323*, 915–919.
16. J. Iwasaki, M. Mochizuki, N. Nagaosa, *Nat. Commun.* **2013**, *4*, 1463.
17. J. Sampaio *et al.*, *Nat. Nanotechnol.* **2013**, *8*, 839–844.
18. T.-E. Park *et al.*, *Phys. Rev. B* **2021**, *103*, 104410.
19. Y. Wu *et al.*, *Nat. Commun.* **2020**, *11*, 3860.
20. Z. Wang *et al.*, *Nat. Commun.* **2018**, *9*, 2516.
21. Z. Yan *et al.*, *Phys. Rev. B* **2022**, *105*, 075423.
22. J. Eom *et al.*, *Nat. Commun.* **2023**, *14*, 5605.
23. S. W. Jang *et al.*, *Phys. Rev. Mater.* **2019**, *3*, 031001.
24. S. Grebenchuk, *Adv. Mater.* **2024**, *36*, 2311949.
25. M.-G. Han *et al.*, *Nano Lett.* **2019**, *19*, 7859–7865.
26. M. Khela *et al.*, *Nat. Commun.* **2023**, *14*, 1378.
27. B. Ding *et al.*, *Nano Lett.* **2020**, *20*, 868–874.
28. R. Pal *et al.*, *npj 2D Mater. Appl.* **2024**, *8*, 30.
29. D. Li *et al.*, *Phys. Rev. B* **2024**, *109*, L220404.
30. R. R. Chowdhury *et al.*, *Sci. Rep.* **2021**, *11*, 14121.
31. G. Zhang *et al.*, *Nat. Commun.* **2022**, *13*, 5067.

32. C. Conner *et al.*, *arXiv* **2024**, arXiv:2411.13721.
33. C. Zhang *et al.*, *Adv. Mater.* **2023**, *35*, 2205967.
34. X. Lv *et al.*, *Nat. Commun.* **2024**, *15*, 3278.
35. Z. Li *et al.*, *Nat. Commun.* **2024**, *15*, 1017.
36. R. Saha *et al.*, *npj Spintronics* **2024**, *2*, 21.
37. Y. Dai *et al.*, *Nat. Commun.* **2024**, *15*, 1129.
38. S. Y. Park *et al.*, *Nano Lett.* **2020**, *20*, 95–102.
39. R. R. Chowdhury *et al.*, *Phys. Rev. Mater.* **2022**, *6*, 014002.
40. H. K. Gweon *et al.*, *Nano Lett.* **2021**, *21*, 1672–1679.
41. R. Z. Zhang *et al.*, *Nano Lett.* **2020**, *20*, 5030–5036.
42. H. Zhang *et al.*, *Nat. Commun.* **2024**, *15*, 3828.
43. T. J. Kim, S. Ryee, M. J. Han, *npj Comput. Mater.* **2022**, *8*, 1.
44. S. Eisebitt *et al.*, *Nature* **2004**, *432*, 885–888.
45. B. Pfau, S. Eisebitt, *RSC Adv.* **2023**, *25*, 254–274.
46. C. Klose *et al.*, *Nature* **2023**, *614*, 256–261.
47. R. Battistelli *et al.*, *Optica* **2024**, *11*, 234–242.
48. Y. Wu *et al.*, *Adv. Mater.* **2022**, *34*, 2110583.
49. C. A. Akosa *et al.*, *Phys. Rev. Appl.* **2019**, *12*, 054032.
50. Y. You *et al.*, *Phys. Rev. B* **2019**, *100*, 134441.
51. Y. Wang *et al.*, *Phys. Rev. B* **2019**, *100*, 024434.
52. D. S. Kathyat, A. Mukherjee, S. Kumar, *Phys. Rev. B* **2020**, *102*, 075106.
53. S. Kumar, A. Majumdar, *Eur. Phys. J. B* **2005**, *46*, 315–323.
54. J.-X. Zhu *et al.*, *Phys. Rev. B* **2016**, *93*, 144404.
55. H. Wu *et al.*, *Phys. Rev. B* **2024**, *109*, 104410.
56. Y. Jiang *et al.*, *Nat. Commun.* **2022**, *13*, 5960.
57. J.-M. Jancu *et al.*, *Phys. Rev. B* **2005**, *72*, 193201.
58. C. M. Moehle *et al.*, *Nano Lett.* **2022**, *22*, 8601–8607.
59. W. Mayer *et al.*, *ACS Appl. Electron. Mater.* **2020**, *2*, 2351–2356.
60. M. B. Hastings, T. A. Loring, *Ann. Phys.* **2011**, *326*, 1699–1759.
61. T. A. Loring, M. B. Hastings, *Europhys. Lett.* **2010**, *92*, 67004.
62. A. Agarwala, V. B. Shenoy, *Phys. Rev. Lett.* **2017**, *118*, 236402.
63. H. Huang, F. Liu, *Phys. Rev. Lett.* **2018**, *121*, 126401.
64. H. Huang, F. Liu, *Phys. Rev. B* **2018**, *98*, 125130.
65. H. D. Rosales *et al.*, *Phys. Rev. B* **2015**, *92*, 214439.
66. D. S. Kathyat *et al.*, *Phys. Rev. B* **2021**, *103*, 035111.
67. L. Powalla *et al.*, *Phys. Rev. B* **2023**, *108*, 214417.
68. L.-M. Kern *et al.*, *Adv. Mater.* **2025**, *37*, 2501250.
69. M. J. Calderón, L. Brey, *Phys. Rev. B* **1998**, *58*, 3286–3292.
70. S. Jin *et al.*, *J. Materiomics* **2025**, *11*, 100865.
71. X. Zhou *et al.*, *Adv. Mater.* **2025**, *37*, e2505977.
72. M. Corasaniti *et al.*, *Phys. Rev. B* **2020**, *102*, 161109.
73. H.-H. Yang *et al.*, *2D Mater.* **2022**, *9*, 025022.
74. A.-A. Haghighirad, *Jpn. J. Appl. Phys.* **2025**, *64*, 040805.
75. D. Rana *et al.*, *Phys. Rev. B* **2023**, *108*, 184419.
76. K. Kim *et al.*, *Nat. Mater.* **2018**, *17*, 794.

# Supplementary Information

## Supplementary Note 1:

**Scaling approach for atomistic simulation of skyrmions.** In atomistic simulations, it is computationally impractical to reproduce the real spatial dimensions of experimental systems. For instance, skyrmions observed experimentally typically have diameters on the order of 100 nm, corresponding to approximately 500 × 500 lattice sites per skyrmion. Simulating a system containing multiple skyrmions would thus require a lattice of about 50,000 × 50,000 sites, far beyond the feasible limit for any Monte Carlo simulation.

To make the problem tractable while retaining the essential physics, we performed simulations on a 120 × 120 lattice and adopted an effectively larger spin-orbit coupling (SOC) strength than that estimated for the real material. The characteristic skyrmion size is known to decrease monotonically with increasing ratio of the SOC strength to the hopping parameter ($\lambda/t$)[1-4]. By increasing $\lambda$, we reduce the characteristic length scale, enabling the formation of smaller skyrmions compatible with our simulation cell. This scaling approach effectively maps the experimental system onto a reduced computational lattice, reproducing the same qualitative skyrmion textures, stability trends, and field-dependent behavior while maintaining computational feasibility.

**Temperature dependence of model parameters.** All model parameters are expressed in units of the hopping parameter $t$, which serves as the fundamental energy scale. In double-exchange systems, the electronic bandwidth is known to undergo strong renormalization with temperature. Consequently, if $t$ is assumed to remain constant, the effective values of all other model parameters become temperature-dependent. This temperature dependence plays a crucial role in reproducing the experimental trends observed in our simulations. At the lowest simulated temperature, the parameters are set to $\lambda$ = 0.2, $A_{\mathrm{u}}$ = 0.10, and $H_z$ = 0.025, corresponding to an experimental magnetic field of -16 mT. At elevated temperatures, these effective parameter values increase due to the reduction of the electronic bandwidth.

**Discretization effects on skyrmion topology.** For an ideal single skyrmion, the quantity integrated over the lattice area should equal unity. Such definitions assume a continuum vector field describing the spin-texture. In numerical simulations, however, the magnetization is sampled only at discrete lattice points, and the continuous integral is replaced by a finite summation over plaquettes. As a result, the computed value may deviate from an exact integer. For example, an ‘almost perfect’ skyrmion might yield a value of 0.9 instead of 1, while a nearly uniform or topologically trivial configuration may produce a small but finite value, such as 0.05. Therefore, this quantity should be regarded as a quantitative measure of how closely a spin texture approximates a topologically non-trivial configuration, rather than as an exact count. Small deviations from integer values naturally arise due to lattice discretization and slight deformations of the spin-texture. Importantly, the value remains nearly constant as long as the topological character of the spin configuration is preserved, even if its shape or size changes significantly.

**Supplementary Note 2:**

**Effective chiral interactions in a centrosymmetric FGT**. FGT crystallizes in the centrosymmetric space group $P6_3/mmc$[5], and therefore does not support a global Dzyaloshinskii-Moriya interaction (DMI), which requires broken inversion symmetry. Consequently, the chiral magnetic interactions responsible for noncollinear spin-textures in FGT cannot be attributed to an intrinsic bulk DMI.

In realistic samples, however, inversion symmetry can be broken locally. In exfoliated thin flakes, local inversion symmetry breaking may arise at surfaces and interfaces, as well as from structural inhomogeneities including defects, strain, or stacking variations. While these effects do not imply a globally non-centrosymmetric phase, they can generate effective chiral interactions at the microscopic level.

To capture these effects, we employ a Rashba-type SOC in our theoretical framework. Rashba SOC provides a minimal and generic description of systems where inversion symmetry is locally broken, leading to momentum-dependent spin splitting. In itinerant electron systems with strong Hund's coupling, such as FGT, the combination of Rashba SOC and electron-spin coupling gives rise to emergent chiral interactions in the effective spin Hamiltonian, including DMI-like terms. In this approach, these interactions are not introduced phenomenologically as an independent DMI term, but instead arise naturally from the underlying electronic Hamiltonian (Eq. (2), Method in the main text).

Band-structure calculations and angle-resolved photoemission spectroscopy (ARPES) studies of FGT reveal strong SOC effects and unconventional band topology[6]. These observations support the relevance of SOC in the low-energy physics of FGT and motivate an effective description that includes SOC-driven chiral interactions. In this context, the Rashba-type SOC used here should be regarded as a minimal effective theoretical construction.

**Role of magnetic inhomogeneities and emergent spin-textures.** The inhomogeneous spin-textures observed in this work (labyrinth domains, isolated skyrmions, and mixed phases) arise dynamically from the competition between exchange interactions, magnetic anisotropy, SOC, dipolar interactions, and external magnetic fields. These inhomogeneities are therefore intrinsic to the energetics of the system, rather than being imposed by quenched disorder.

**Topological characterization via the Bott index.** To characterize the topology of the electronic states in the presence of spatially varying magnetization, we compute the Bott index (See Method in the main text). This quantity is evaluated from the simulated electronic states corresponding to the real-space spin configurations and is known to be equivalent to the Chern number in a magnetically non-periodic system. The Bott index thus provides a measure of the topology of the electronic degrees of freedom in systems with inhomogeneous magnetic textures.

## Supplementary Figures

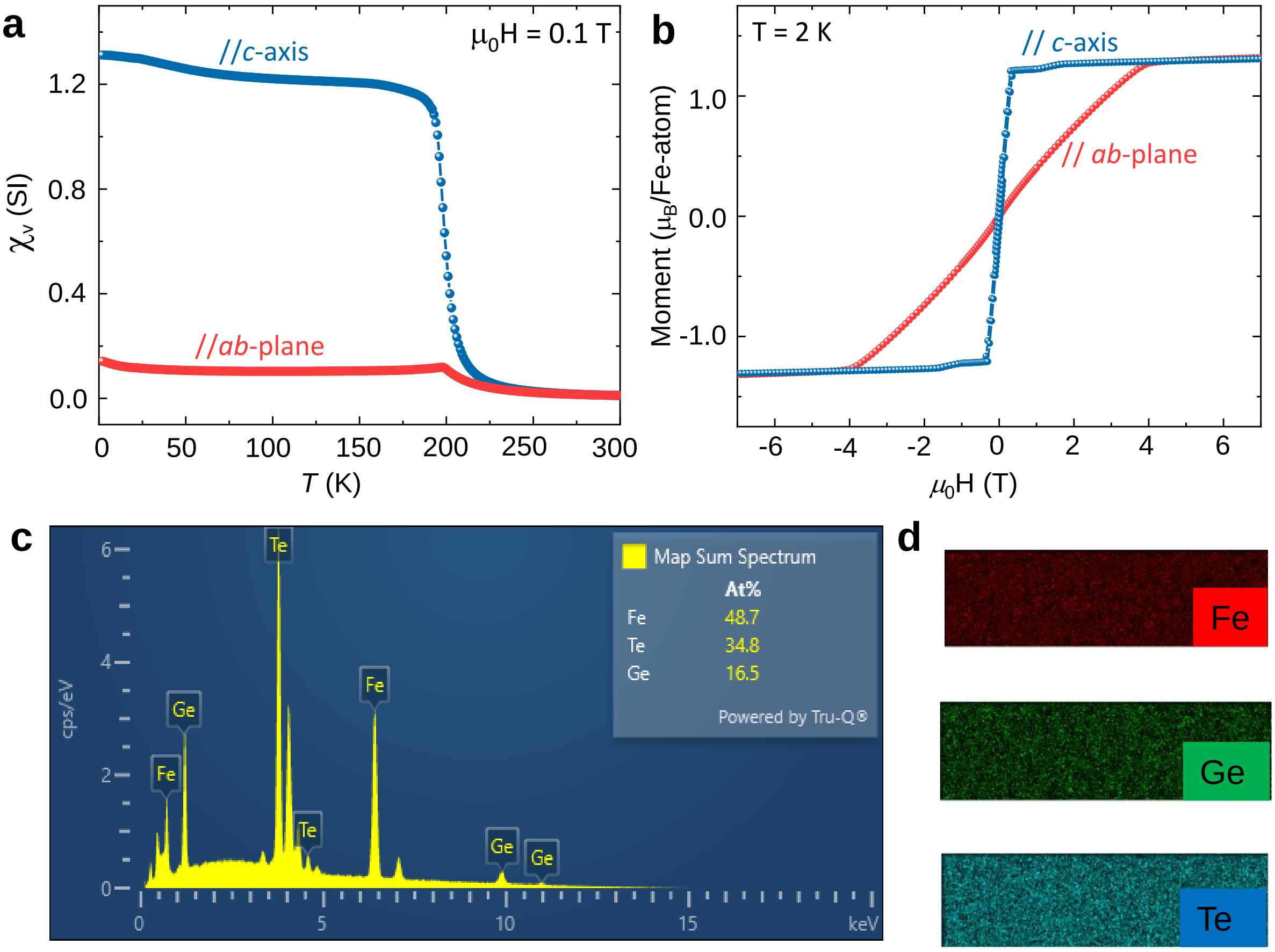

**Supplementary Fig. 1| Physical characterization of FGT single crystals: composition and magnetism. a,** Temperature-dependent magnetic volume susceptibility $\chi_V(T)$ measured during field-cooled warming at $H = 0.1$ T with the field applied along the *ab*-plane ($H \parallel ab$) and along the *c*-axis ($H \parallel c$). The $T_C \approx 205$ K is consistent with a near-stoichiometric $Fe_3GeTe_2$ composition. **b,** Isothermal magnetization *M-H* hysteresis curves at 2 K for $H \parallel ab$ and $H \parallel c$, showing strong uniaxial anisotropy with the *c*-axis as the easy axis. **c,** Surface elemental mapping confirming compositional uniformity. **d,** Display corresponding energy dispersive spectra (EDS) elemental maps over the same region. The individual elemental maps of Fe, Ge, and Te confirm the spatial distribution of each species.

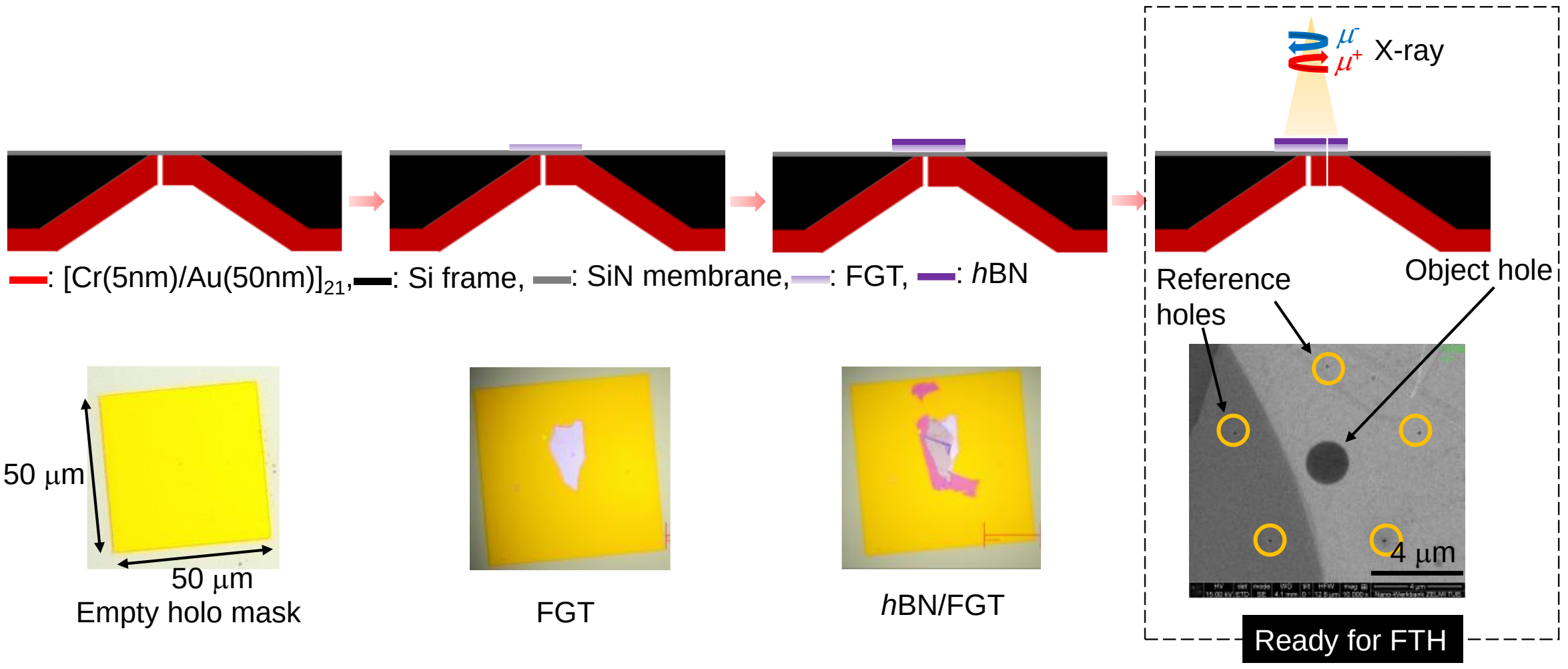

**Supplementary Fig. 2| Sample preparation workflow for synchrotron-based Fourier transform holography (FTH). Upper row:** Schematic illustrates the sequential sample-preparation steps enabling high-spatial-resolution magnetic imaging by FTH. **Lower row:** Corresponding optical micrographs of FGT thin flake at each step, aligned with the schematic above. The FGT flake is encapsulated with ~20 nm of hexagonal boron nitride (*h*BN) to prevent oxidation during handling and measurements. The dotted rectangle displays an SEM image of the final sample configuration, ready for FTH. The object hole (defining the field of view) and the reference holes, both critical for holographic reconstruction, were milled by focused ion beam (FIB).

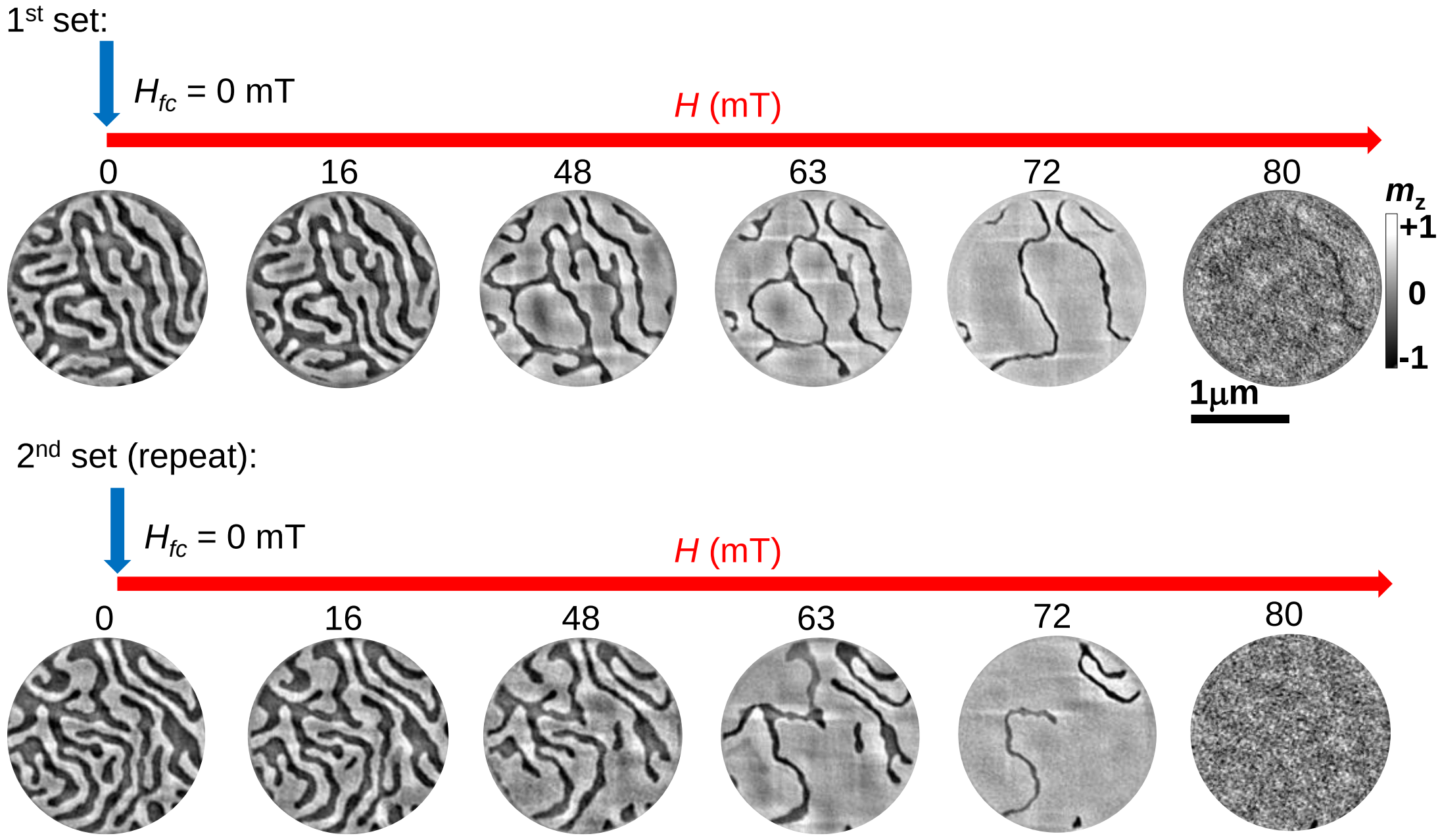

**Supplementary Fig. 3| Absence of pinning in the spin-texture of FGT.** Spin-textures in a thin FGT flake under applied magnetic fields at 130 K, after repeated field cooling through $T_C$ with $H_{fc} = 0$ mT. Although magnetic nanodomains appear at different locations within the same region of the flake, they consistently exhibit an overall labyrinthine pattern, indicating the absence of strong pinning. The evolution of spin-textures remains qualitatively similar across multiple field-cooling cycles.

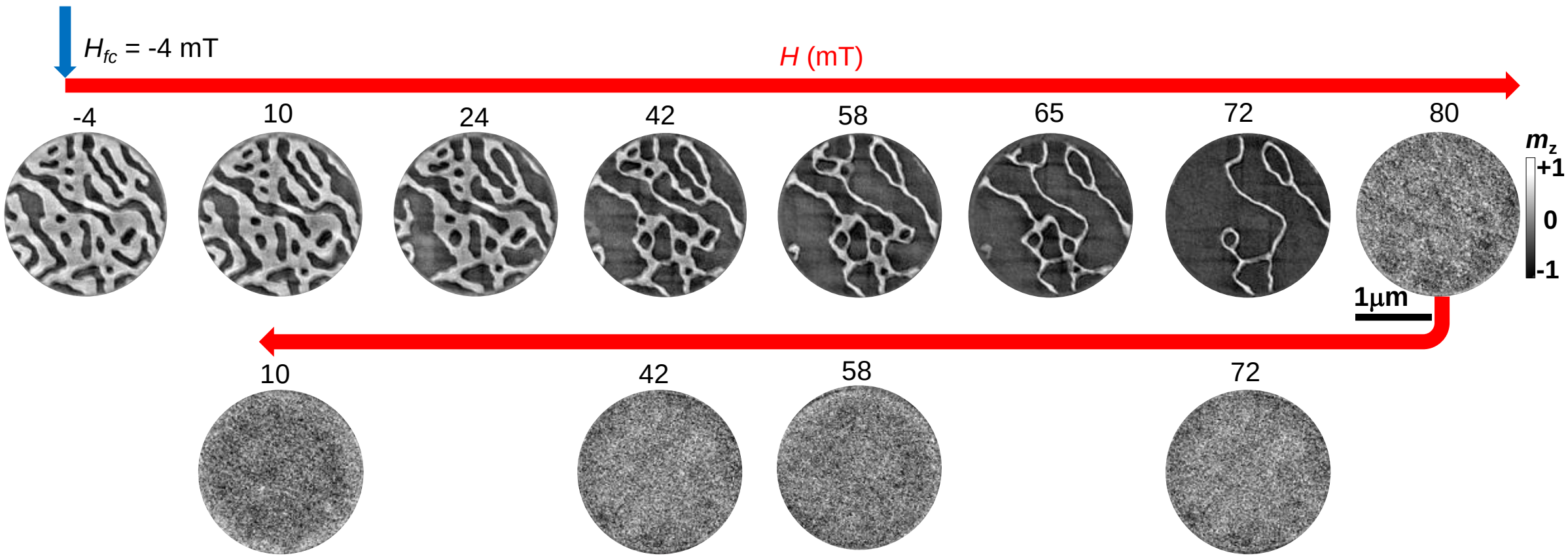

**Supplementary Fig. 4| Observation of skyrmion bags in FGT.** Spin-textures in a thin FGT flake under increasing magnetic field $H$ at 130 K after field cooling through $T_C$ at $H_{fc}$ = -4 mT. The initial spin-configuration at -4 mT is a mixed labyrinth-skyrmion phase. As the field increases, skyrmion bag structures emerge at intermediate fields of 42, 58, 65, and 72 mT. With further field increase, the system transitions to a uniform magnetic domain state, which persists after field reversal.

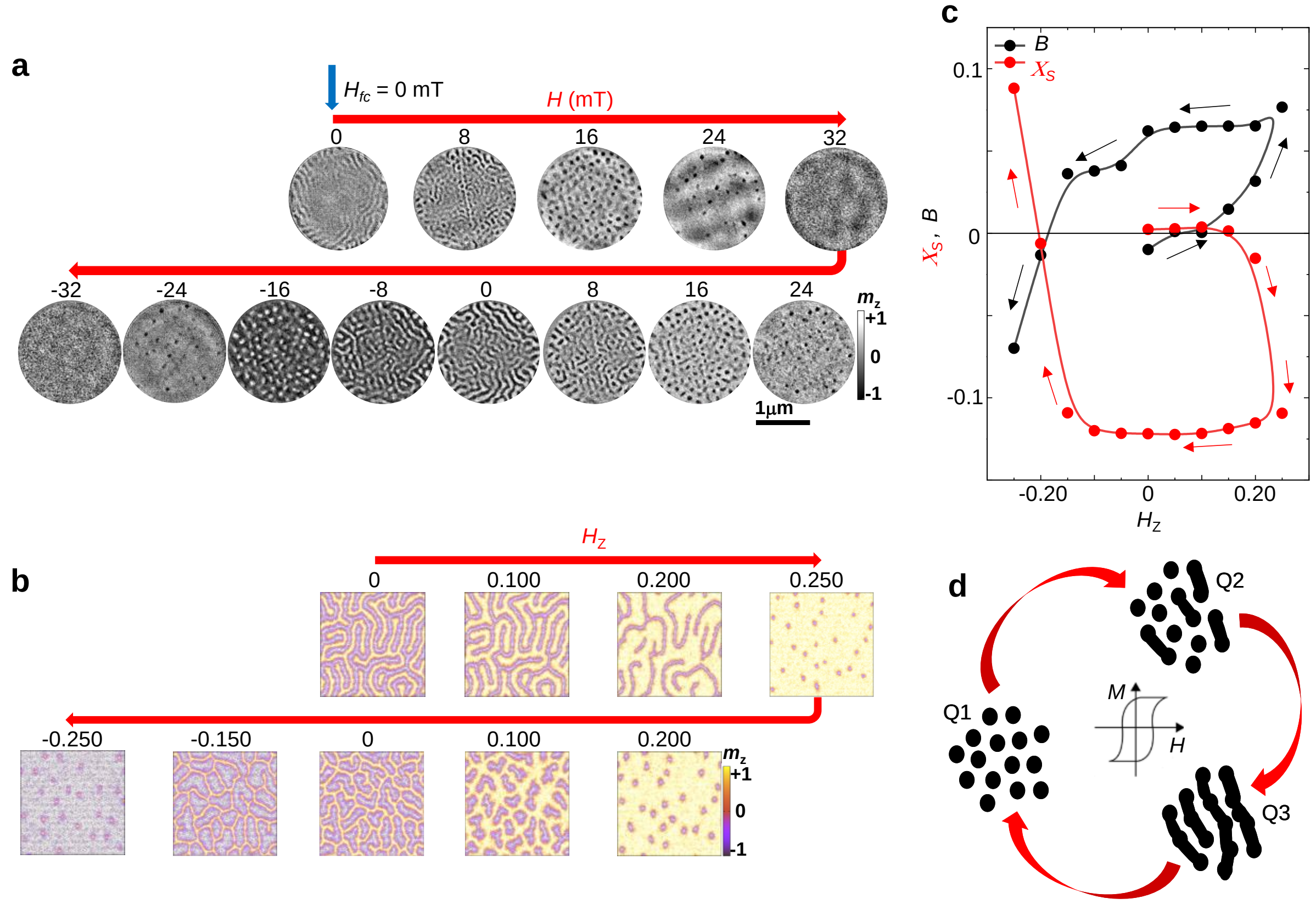

**Supplementary Fig. 5| Topological transitions and unprotected states near $T_C$ (194 K). a,** Experimental real-space spin-textures in an FGT thin flake during magnetic-field cycling at 194 K after field cooling through $T_C$ at $H_{fc}$ = 0 mT. **b,** Theoretical real-space spin-textures from simulations on a $N = 120^2$ system, after field cooling at $H_Z$ = -0.050 ($T$ = 0.07, $\lambda$ = 0.3, $A_u$ = 0.12). **c,** Variation of skyrmion density $\chi_s$ and Bott index $B$ as a function of $H_Z$, showing a clear topological transition characterized by a zero-line crossing in both quantities. **d,** Schematic depicting topological transitions and the emergence of unprotected states during field cycling.

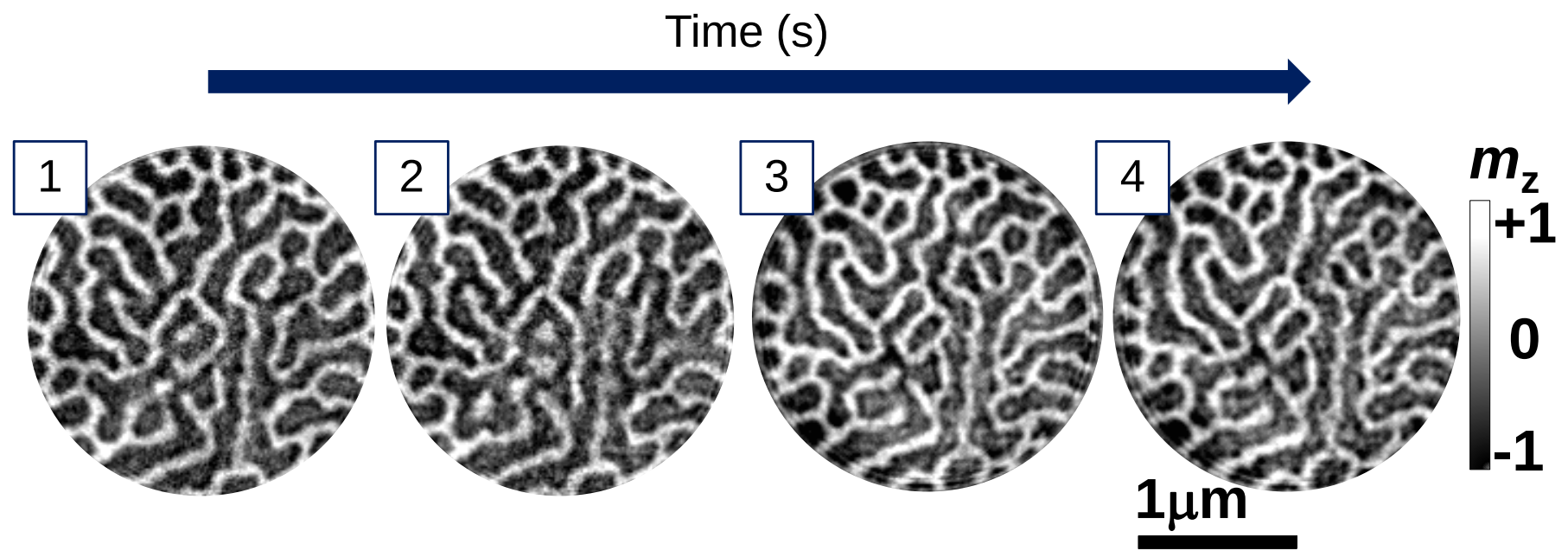

**Supplementary Fig. 6| Temporal dynamics of spin-textures near $T_C$ (194 K): a case study of -8 mT data of Fig. 5.** Domain configurations of four sequential time segments (1-4) captured at -8 mT during data acquisition, illustrating time-resolved evolution and fluctuations of magnetic domain structures in FGT thin flake. These frames highlight the dynamic character of spin-textures near $T_C$.

## References

1. P. Sahu, B. R. K. Nanda, S. Satpathy, *Phys. Rev. B* **2022**, *106*, 224403.
2. H. Wu, *et al.*, *Commun. Phys.* **2021**, *4*, 210.
3. X. R. Wang, X. C. Hu, H. T. Wu, *Commun. Phys.* **2021**, *4*, 142.
4. N. Nagaosa, Y. Tokura, *Nat. Nanotechnol.* **2013**, *8*, 899.
5. H. Deiseroth, K. Aleksandrov, C. Reiner, L. Kienle, R. Kremer, *Eur. J. Inorg. Chem.* **2006**, *2006*, 1561.
6. K. Kim, J. Seo, E. Lee, K.-T. Ko, B. S. Kim, B. G. Jang, J. M. Ok, J. Lee, Y. J. Jo, W. Kang, J. H. Shim, C. Kim, H. W. Yeom, B. Min, B.-J. Yang, J. S. Kim, *Nat. Mater.* **2018**, *17*, 794.